\documentclass[prb,twocolumn]{revtex4-1} 
\usepackage{graphicx}
\usepackage{dcolumn}
\usepackage{bm}

\begin{document}

\title{The twin paradox in relativity revisited}
 \author{Vasant Natarajan}
 \email{vasant@physics.iisc.ernet.in}
 \homepage{www.physics.iisc.ernet.in/~vasant}
 \affiliation{Department of Physics, Indian Institute of
 Science, Bangalore 560\,012, INDIA}

\begin{abstract}
The standard resolution of the twin paradox in relativity states that the age of the inertial twin `jumps' when the traveling twin undergoes his turn-around acceleration. This resolution is based on the use of the equivalent gravitational shift in the frame of the accelerating twin. We show that this is an incorrect and untested use of the equivalence principle. We also propose an unambiguous test of the standard resolution based on the Pound-Rebka experiment.
\end{abstract}

\maketitle

The twin paradox in relativity has a history almost
as long as the theory of relativity itself. It was
originally proposed by Einstein as a {\em gedanken}
experiment to highlight the fact that an observer sees a
moving clock going at a slower rate than a clock at rest.
Because this is an entirely symmetric effect (in that each
clock sees the other clock ticking slower), the paradox
arises when two identical synchronized clocks are
temporarily separated into different Lorentz frames and
then brought back together. Applied to twins, the paradox
starts with identical twins on Earth. One of the twins then
accelerates away on a rocket, moves away from Earth at a
constant velocity $u$ for a time $T/2$, fires rockets to
accelerate again so that his velocity changes to $-u$,
moves at the constant velocity $u$ towards the Earth for a
time $T/2$, and decelerates to a stop on reaching the Earth
again. The acceleration times are assumed to be negligible
compared to $T$. The paradox arises because both twins
observe the other to be aging slower during the period of
uniform relative motion. Are they the same age when they
meet again or is one of them younger, and if so, which one
and by how much?

Over the years, the paradox has been discussed extensively
in many books and articles. The American Journal of Physics has carried a
large number of articles on the paradox,
\cite{GRE72,GIG79,BOU89} highlighting the difficulty in
conveying this concept to first-time students of
relativity. It still remains one of the most puzzling
aspects of the theory of relativity.

In the standard resolution, as presented in many textbooks
on relativity, both twins conclude that the
traveling twin who {\em accelerated} is younger. The
argument proceeds as follows. The Earth-bound twin always
remains in an inertial frame and therefore his observation
that the other twin is aging slower is {\em correct}. On
the other hand, the rocket-bound twin sees his brother age
slowly during the time when the relative velocity is
constant, but sees a {\em sudden jump} in his brother's age
during the short acceleration phase when he is not in an
inertial frame. Thus, the change of inertial frames results
in a jump in age. The use of the word sudden, which is standard terminology in these discussions, is not meant to imply that the change is discontinuous but only to mean that the change happens during the relatively short acceleration phase.

Since the paradox involves an acceleration phase for at
least one of the twins, it is useful to bring in some of
the framework of general relativity to analyze this
problem. A mathematical analysis along these lines is presented by R.\ C.\ Tolman in his book
{\em Relativity Thermodynamics and Cosmology}.\cite{TOL87} His main argument is that the jump in age seen by the
rocket-bound twin during the acceleration phase is due to
the {\em equivalent gravitational shift} between clocks
placed at different points in a uniform gravitational
field. By the equivalence principle, the acceleration can
be viewed equally well as arising due to the turning on of
a uniform gravitation field. If the distance between the
twins is $D$, the acceleration of the rocket is $g$, and
the duration of acceleration (in the rocket frame) is
$\tau_R$, then in this time the accelerated twin will
perceive his Earth-bound brother to age by an amount
\begin{equation}
\tau_E \approx \tau_R \left( 1 + \frac{g D}{c^2} \right) \,
,
\end{equation}
where the $\approx$ sign indicates that the relation is
correct to first order. This change can be very large
compared to $\tau_R$ for large values of $D$. In other words, the short acceleration time in the rocket frame is equal to a large time in the Earth frame.

Using this idea, we can calculate the ages of the twins
from both viewpoints. For the Earth-bound twin, if the time
of uniform relative motion is $T_E$ and the three
acceleration phases have negligible duration, then his age
has increased by $T_E$, while his brother's age $T_R$ is
shorter and related to $T_E$ by:
\begin{equation}
T_E = \frac{T_R}{\sqrt{1-u^2/c^2}} = T_R \left(1 + u^2/2c^2
+ \ldots \right) .
\label{e1}
\end{equation}
From the viewpoint of the rocket-bound twin, the equivalent
gravitational shifts during the three acceleration phases
should also be taken into account. During the initial
acceleration away from Earth and the final deceleration
upon returning to Earth, the twins are almost at the same
gravitational potential and the differential shift is
negligible. However, during the intermediate turn-around
acceleration, the twins are widely separated and the shift
is large. If the travel time (in the rocket frame) before
turn around is $T_R/2$, then their separation is $ D = u
T_R /2 $. Since the velocity changes by $2u$ in a time
$\tau_R$, the acceleration is $ g = 2u / \tau_R $.
Therefore, the age of the Earth-bound twin as seen by his
brother is:
\begin{equation}
T_E = T_R \sqrt{1-u^2/c^2} + \tau_R \left( 1+
\frac{T_R}{\tau_R} \frac{u^2}{c^2} \right) .
\label{gr1}
\end{equation}
Neglecting the term proportional to $\tau_R$, this is the
same relation as Eq.\ (2), and thus both twins agree that
the rocket-bound twin is younger, and by the same amount
(at least to leading order).

We thus see that the successful resolution of the paradox from the viewpoint of the
traveling twin rests on the statement that the accelerated
clock will see the far-away clock go at a slower rate due
to the equivalent gravitational shift predicted by general relativity. In the words of Tolman:\cite{TOL87} \\
{\em The solution thus provided for the well-known clock
paradox of the special theory $\ldots$ has been made
possible by the adoption of the general theory of
relativity}.

The above analysis also shows why the following statement from Ref.\ 3
is incorrect:\\
{\em It has often been pointed out that while the acceleration of one twin is the key to the resolution of the paradox, it is wrong to suppose that reduced aging is a direct result of acceleration. The age difference of the twins is proportional to the length of the trip while the period of acceleration is determined only by how long it takes to turn around and is independent of the length of the trip and, hence, the final age difference of the twins.}\\
Eq.\ \ref{gr1} shows clearly that the age difference is indeed proportional to the length of the trip ($T_R$) while the period of acceleration ($\tau_R$) cancels out. In other words, changing the period of acceleration changes the value of $g$ exactly by the amount required to produce an age difference proportional to $T_R$.

Some authors argue that there is no need to bring in the
framework of general relativity to resolve the paradox.
However, even when the analysis stays within the domain of special relativity, these authors conclude that there is a sudden jump in age of the far-away
twin during the acceleration phase. To quote from the
classic textbook {\em Spacetime Physics: Introduction to
Special Relativity} by Taylor and Wheeler (Ref.\ 5,
page 130):\\
{\em This `jump' $\ldots$ is the result of the traveler
changing frames. And notice that the traveler is unique in
the experience of changing frames; only the traveler
suffers the terrible jolt of reversing direction of motion.
In contrast, the Earth observer stays relaxed and
comfortable in the same frame during the astronaut's entire
trip.}

Despite the above arguments, Tolman's general relativistic analysis has two features that will prove useful to us:
\begin{list}{}{}
\item[(i)] The equivalent gravitational shift when the two clocks are nearby is negligible, so that the two observers agree on the duration of the acceleration times in the first and third phases. In other words, a local inertial observer can calculate the acceleration time in the non-inertial frame.

\item[(ii)] The equivalent gravitational shift when the two clocks are widely separated provides a way of quantitatively calculating the `jump' due to the change in frames.

\end{list}

Furthermore, Tolman's analysis shows that it is not the acceleration (or the terrible jolt of changing frames) {\em per se} but only the acceleration
which happens when the twins are separated that is
important for the resolution  of the paradox. In
fact, this shows that a further simplification is possible. The accelerating twin need not really complete the return
journey. It is sufficient if he just decelerates to come to
rest with respect to his far-away twin, since their
simultaneous ages can now be compared without ambiguity,
although they will be at different locations. The initial acceleration which
both twins experienced (but at the same location) is not
relevant.

We next consider a case discussed widely in the American Journal of Physics, \cite{BOU89} namely one in which both twins experience a {\em symmetric} acceleration phase but one that still results in asymmetric aging because they are spatially separated. Consider twins who are moved to widely different locations in the
Earth frame after birth. Then they get on to spaceships and accelerate identically
by firing identical rockets. After the acceleration phase,
they are both moving in the same direction with identical
velocity of $u$. Thus they are at rest with respect to each
other and their (simultaneous) ages can be compared without ambiguity. There is no doubting the fact that they have gone
through identical experiences with respect to the
acceleration phase. However, according to the earlier
equivalence-principle analysis, the twin on the left would see the one on the right to be older after the
acceleration phase. The only asymmetry is that the direction of acceleration is towards the other twin, so that the
left twin would be older if they both accelerated to the
left instead. In the language of {\em Spacetime Physics} (Ref.\ 5,
page 117), the twins are both jumping on to a moving frame at different locations.

The analysis using the equivalence principle proceeds as follows. Let the duration of the uniform acceleration phase in the frame of the left twin be $\tau_L$. Then the acceleration is $g = u/ \tau_L$, so that the equivalent duration for the right twin at a distance $x_0$ away is
\begin{equation}
\tau_R = \tau_L \left( 1 + \frac{u x_0}{\tau_L c^2} \right) .
\end{equation}
To give some concrete numerical values, let us take $\tau_L$ to be 1 yr and $u x_0 / c^2$ also to be 1 yr. Thus the left twin concludes that his age after the acceleration phase has increased by 1 yr while that of his twin has increased by 2~yrs. From the point of view of the right twin, the acceleration is $g = u/ \tau_R$, but now he is at the higher gravitational potential so that
\begin{equation}
\tau_L = \tau_R \left( 1 + \frac{u x_0}{\tau_R c^2} \right) ^{-1} .
\end{equation}
Hence he will reach {\em the same conclusion} that the twin on the left is younger. But, since the duration of the acceleration phase {\em in the rest frame} is the same for both twins, $\tau_R$ is 1~yr and the age of the twin on the left has increased only by 0.5~yr. In other words, while the twins will agree that the twin on the right is older, they will disagree on the amounts of aging. The fact that the acceleration phase lasts for the same (proper) time seems correct because it depends on the amount of fuel and the rate of burning, which is taken to be identical for the two rockets. This is also consistent with point (i) of Tolman's argument, namely that a locally inertial observer can verify that the duration of the acceleration phases in the non-inertial frame are identical.

The above analysis shows that the use of the equivalence principle in a standard variation of the twin paradox, namely the case of the identically accelerated twins, results in an inconsistent answer. It therefore begs the question as to whether the equivalence principle, which is the bedrock of general relativity and undoubtedly correct, is being used properly in this case. Before we answer the question, let us consider the statement of the principle from Weinberg's book on Relativity and Cosmology:\cite{WEI72} \\
{\em at every space-time point in an arbitrary
gravitational field it is possible to choose a ``locally
inertial coordinate system'' such that, within a
sufficiently small region of the point in question, the
laws of nature take the same form as in unaccelerated
Cartesian coordinate systems in the absence of
gravitation.} \\
The key word in the above statement is {\em locally}, defining a region which is ``sufficiently small'' so that the gravitational field is constant and tidal effects can be ignored. On the other hand, the gravitational redshift is an explicitly {\em nonlocal} phenomenon, relating to the relative rates of two clocks placed at different potentials in a gravitational field.

Therefore, the use of the equivalent gravitational shift in an accelerated frame requires adequate care. To see how to derive this, we follow the procedure given by Tolman in his book.\cite{TOL87} Consider two identical clocks, placed left and right and separated by a distance $D$, in a frame being uniformly accelerated to the right at a rate $g$. Let the first clock emit a photon at time $t=0$. Traveling at $c$, the photon reaches the second clock after a time $D/c$. In this time, the second clock has picked up an additional speed of $gD/c$, so that the first-order Doppler effect yields for the relative rates
\[
\frac{\tau_R}{\tau_L} =  1 + \frac{g D}{c^2}  \, ,
\]
exactly as expected for the gravitational redshift in a uniform gravitational field. Thus, the physical basis for the shift in an accelerated frame is the normal Doppler effect. But this derivation is valid in a frame undergoing constant acceleration that lasts forever, or at least as long as it takes for a photon to reach from one clock to the other. This is explicitly not the case in the twin-paradox experiment, where the duration of the acceleration phase is negligibly small and certainly not long enough for a photon to cover the distance between the twins. The increased relative velocity can at most be $u$ (depending on the instant at which the photon was emitted), and definitely not $gD/c$. This shows clearly why the equivalent shift cannot be applied to the short acceleration phase of the twin-paradox kind.

This also shows that the use of the equivalence principle in the standard resolution to the twin paradox {\em remains untested}. We therefore propose a straightforward and unambiguous test of the jump in age predicted by the twin-paradox effect with the following experiment. We propose a repeat of the Pound-Rebka measurement of the gravitational redshift using recoilless gamma rays emitted by iron nuclei,\cite{POR60b} but this time with the source and absorber at the same level and {\em with the absorber accelerated towards the source at a constant rate $g$}. Since the source and absorber are at the same gravitational potential, there will be no frequency shift due to the gravitational redshift. However, the twin-paradox effect predicts a shift of $gD/c^2$, where $D$ is the distance between the two. This is an exact reproduction of the twin-paradox experiment, and should provide conclusive results one way or the other.

There are several attractive features of the proposed
experiment. First is that, since the sign of the shift depends on the direction of acceleration, several sources of systematic error can be eliminated by looking for a
difference between acceleration to the right and
acceleration to the left. Secondly, the shift depends linearly
on the distance $D$. Therefore, one can increase the
sensitivity of the experiment by increasing $D$. Note that
the original Pound-Rebka experiment was done with a height
difference of 74 ft, while on the ground the distance could
be increased by a factor of 2 or more. Finally, there will be a small shift from the first-order Doppler effect depending on the exact velocity of the source at the time of emission, but there are two experimental handles to address this error: (i) the shift will be limited in magnitude by the final velocity of the absorber which can be made small, and (ii) it will be independent of $D$.

Let us also consider here an experiment \cite{POR60a} that is often quoted as
an experimental verification of asymmetric aging in the
twin paradox (for example in Ref.\ 5,
page 134). This is also an experiment
by Pound and Rebka, but where the frequency of the recoilless gamma rays were studied as a
function of temperature. This experiment was done primarily
to check for sources of systematic error in their
gravitational redshift experiment. They did find a shift in
the frequency of the gamma ray as the temperature was
increased. However, this is not a test of the twin-paradox
effect. On the contrary, this is only a measurement of the
second-order Doppler effect, because the increased
temperature results in an increased average value of $u^2$.
Indeed, if we could transform to the frame of the
oscillating iron nuclei, the laboratory clocks would appear
to go slower! This is a manifestation of time dilation
similar to the {\em apparently} long lifetime of particles
decaying in flight as observed from a stationary frame. Nobody will argue that particles in flight age slower because  of this apparent dilation.

We conclude by discussing what, in the author's opinion, is the logical fallacy in the standard resolution to the twin paradox. The fallacy is to ascribe a {\em real} change to the {\em apparent} effect of time dilation between relatively moving frames. This is akin to saying that the well-known phenomenon of Lorentz contraction is real and causes a real change in the length of a moving rod. The length of a rod is determined by interatomic forces and is not dependent on its motion with respect to some arbitrary frame, while the Lorentz contraction is an apparent effect due to the relativity of simultaneity. Similarly, the age of a person (or clock) is determined by physical processes that control decay rates. Moving one person temporarily to another Lorentz frame cannot change this rate. Indeed, it can be shown that the differential aging in the twin-paradox effect arises from the fact that the distance to the turn-around point appears Lorentz contracted in the Rocket-bound frame and hence takes a shorter time to travel. This leads to the unlikely conclusion that travel to a distant star is possible if we travel close to the speed of light. The nearest star (outside the solar system) is 4.5 light years away, so that an astronaut traveling at $0.9c$ would take 5 yrs (in the Earth frame) to get there. But the twin-paradox effect predicts that the astronaut would have aged only by $5 \sqrt{1-0.9^2}=2.18$ yrs (Eq.\ \ref{e1}). From the Lorentz contraction point of view, the distance to the star appears shorter by the same factor, and therefore takes a shorter time to travel. This suggests that the distance to any faraway object can be made arbitrarily small by traveling sufficiently close to the speed of light. Not very likely!

It is also important to realize that the gravitational redshift is a real effect on the relative rates of clocks placed at different gravitational potentials. The curvature of spacetime in the presence of a gravitational field has a physical effect on natural processes so that there is a difference in clock rates when one clock is compared to the other, though no local measurements on one clock will show any change because all processes are affected the same way. Therefore, if one twin climbs up to the top of a mountain (i.e.\ to a higher gravitational potential) while his brother stays at the base, lives atop for one year, and then comes back down to rejoin his brother, he will find that his age is slightly higher. But this is because the curved spacetime near the Earth's surface causes a real change in the relative heart rates of the two brothers. But mere acceleration in flat spacetime for a short time (short enough so that the normal Doppler effect can be neglected) can not cause such a change, as required in the twin-paradox effect. We have therefore proposed a simple experiment to test this hypothesis.

\begin{acknowledgments}
The author thanks Krishna Parat for introducing him to the twin paradox and Wolfgang Ketterle for useful discussions.
\end{acknowledgments}


\begin{thebibliography}{1}
\newcommand{\enquote}[1]{``#1''}

\bibitem{GRE72}
D.~M. Greenberger, \enquote{The reality of the twin paradox effect,} Am. J.
  Phys. \textbf{40}, 750--754 (1972).

\bibitem{GIG79}
C.~Giannoni and O.~Gron, \enquote{Rigidly connected accelerated clocks,} Am. J.
  Phys. \textbf{47}, 431--435 (1979).

\bibitem{BOU89}
S.~P. Boughn, \enquote{The case of the identically accelerated twins,} Am. J.
  Phys. \textbf{57}, 791--793 (1989).

\bibitem{TOL87}
R.~C. Tolman, \emph{Relativity Thermodynamics and Cosmology} (Dover
  Publications Inc., New York, 1987).

\bibitem{TAW92}
E.~Taylor and J.~A. Wheeler, \emph{Spacetime Physics: Introduction to Special
  Relativity} (W. H. Freeman and Company, New York, 1992), 2nd ed.

\bibitem{WEI72}
S.~Weinberg, \emph{Gravitation and cosmology: Principles and applications of
  the general theory of relativity} (John Wiley and Sons, New York).

\bibitem{POR60b}
R.~V. Pound and J.~G.~V.~Rebka, \enquote{Apparent weight of photons,} Phys.
  Rev. Lett. \textbf{4}, 337--341 (1960).

\bibitem{POR60a}
R.~V. Pound and J.~G.~V.~Rebka, \enquote{Variation with temperature of the
  energy of recoil-free gamma rays from solids,} Phys. Rev. Lett. \textbf{4},
  274--275 (1960).

\end{thebibliography}

\end{document}